\documentclass[11pt,twoside]{article}  
\usepackage{adassconf}
\usepackage{asp2006}

\begin{document}   

%
%
%

\paperID{P4.21 }

%
%
%
%

\title{Recalibration of Data in the VDFS Science Archives}
\titlemark{Recalibration in VDFS}

%
%
%

\author{Nicholas Cross, Nigel Hambly, Ross Collins, Mike Read and Eckhard Sutorius}
\affil{Scottish Universities Physics Alliance, Institute for Astronomy, University of Edinburgh, Blackford Hill, Edinburgh, EH9 3HJ, U.K.}

%
%

\contact{Nicholas Cross }
\email{njc@roe.ac.uk}

%
%
%
%
%

\paindex{Cross, N.}
\aindex{Hambly, N.}
\aindex{Collins, R.}
\aindex{Read, M.}
\aindex{Sutorius, E.}

%
%

\authormark{Cross et al. }

%
%



\begin{abstract}
The VDFS comprises the system to pipeline process and archive the data from
infrared observations taken by both the WFCAM instrument on UKIRT and the
forthcoming VISTA telescope. These include the largest near-IR surveys to
date, such as UKIDSS, which produce terabyte sized catalogues of over 10$^9$
rows. Such large data volumes present a performance challenge when the
catalogue data, stored in a relational database, require many iterations of
astrometric and photometric recalibration. Here we present the VDFS
recalibration solution that will be employed in the WSA from the forthcoming
UKIDSS Data Release 4 and VSA from its inception.
\end{abstract}

%
%

\vspace{-10mm}
\section{Introduction}
The VISTA Data Flow System (VDFS) is designed to pipeline process and
archive the data from infrared observations taken by both the UK Infrared
Telescope's (UKIRT) Wide-Field Camera (WFCAM) and the forthcoming Visible
and Infrared Survey Telescope (VISTA). These are currently the fastest
near-IR survey instruments and will remain so for years to come, producing
terabytes of catalogue data each year. These data are stored in the WFCAM 
and VISTA Science Archives (WSA
\footnote{http://surveys.roe.ac.uk/wsa/pre/index.html} and
VSA respectively) designed and maintained by the Wide-Field Astronomy Unit
(WFAU) in Edinburgh. The archives consist of relational databases of
catalogue data and associated metadata, together with a data store of all
the individual image and catalogue FITS files. The design of the existing
WSA is described in detail by Hambly et al. (2008). 

Calibration of WFCAM and VISTA data is performed at the pipeline 
processing stage by the Cambridge Astronomy Survey Unit (CASU, Hodgkin 
et al. 2008).  Improvements in the calibration have also been calculated 
by CASU in conjunction with a Calibration Working Group containing members 
from CASU, WFAU and UKIRT Infra-Red Deep Sky Survey (UKIDSS). Both the 
astrometric and the photometric calibration are calculated through 
comparison to the Two Micron All Sky Survey (2MASS, Skrutskie et al. 2006), 
which covers $99.998\%$ of the sky resulting in a very uniform calibration
accuracy. 

However, it is a complicated task to convert from the raw photometric and 
astrometric quantities measured by the source extractor to calibrated 
quantities, removing all the instrumentational effects. Subtle effects are
often only measured after significant amounts of data have been taken. 
The equations converting raw data to calibrated quantities will need to be
adjusted to account for these effects thus reducing the error on these 
measurements. 

In addition to improving the calibration of new data, existing data in the
archive will require recalibration, and this process will happen many times
throughout an archive's history. The WSA catalogues increase at a rate of 
$\sim1$TB/year and the VSA catalogues will increase at a rate of 
$\sim3$TB/year. After five years the VSA will contain 15TB of catalogue 
data, a factor of ten larger than the amount of data that
was last recalibrated for UKIDSS Data Release 3. 

\vspace{-5mm}
\section{Recalibration Procedure}

Recalibration of archived data can take many forms. For example, UKIDSS
DR2 required a change in the photometric zero-point measurements
for each observation. In earlier releases, all 2MASS stars were used for
calibration (see Dye et al. 2006 for the equations). For UKIDSS DR2, 
only a colour-cut subset of these stars were used, and a
Galactic extinction term was incorporated into the calibration equation
(Warren et al. 2007). 

\begin{figure}
\plotone{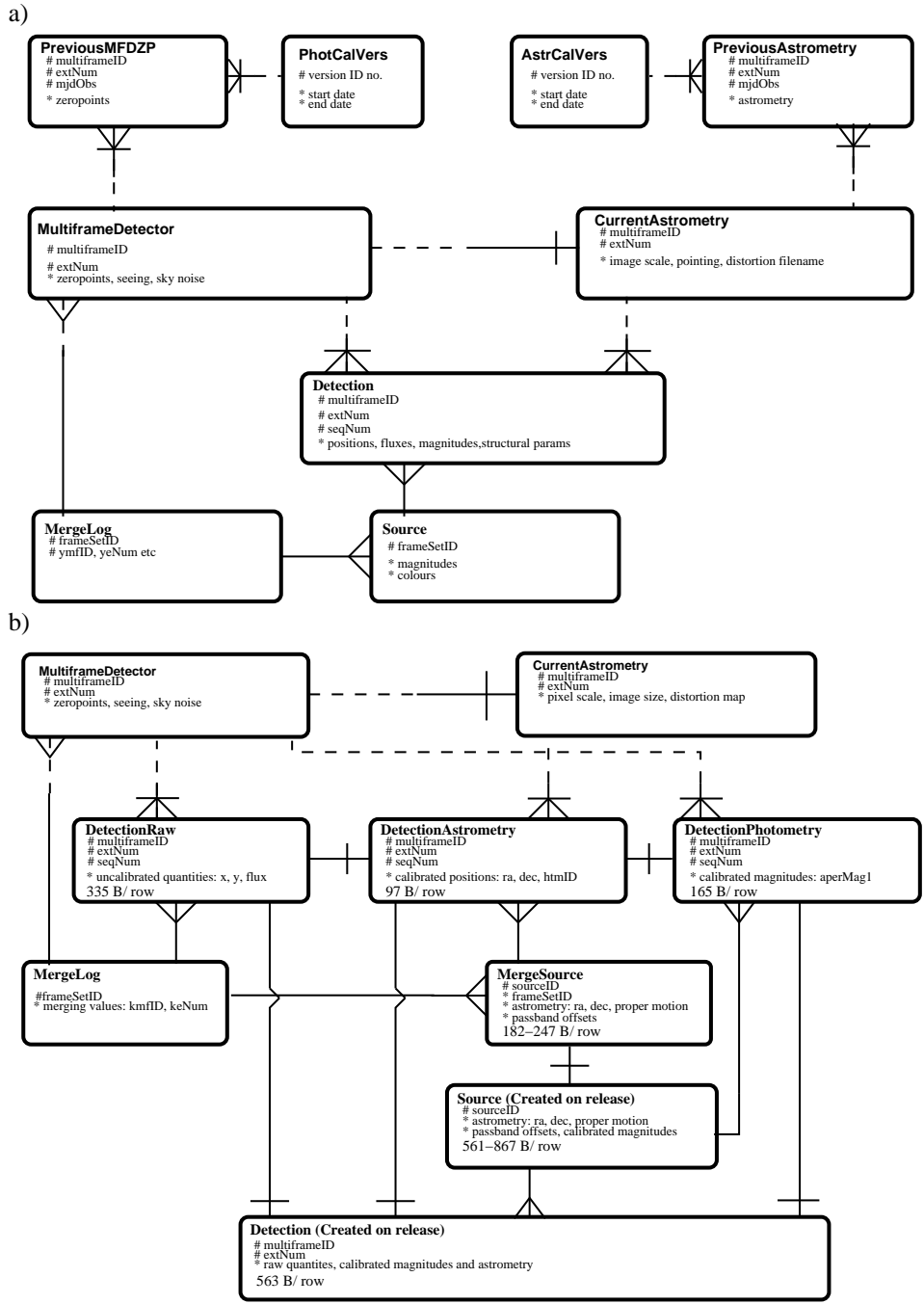}
\caption{\it Entity Relationship Models (ERM) for the current 
(a) and future (b) recalibration schemas. Each box represents a 
database table showing in schematic form the relationship between the 
attributes of the connected tables. A one-to-many relationship, where a 
single entry in one table is related to multiple entries in another, is 
represented by the `crows-foot' connecting the boxes. Dotted lines 
indicate optional as opposed to mandatory relationships and a perpendicular 
line indicates that the tables share a primary key. Hambly et al. (2008) 
provides a full description of the relational model for the WSA.}
\label{fig:ERMs}
\end{figure}

Applying such a recalibration to the archive requires
that both the metadata in the database and FITS file headers be updated 
with the new zero-point values. Also, the catalogue data
quantities that are derived from the calibration equations and stored in the
database detection tables then require adjustment if the change in
zero-point is significant ($|\Delta\,ZP|>5\times10^{-4}$). The old metadata 
values are maintained both in a new HISTORY card in the FITS file headers 
and in a database table, \verb+PreviousMFDZP+, that is designed to store 
only the metadata of earlier calibrations. A history of archive 
recalibration events is maintained in the \verb+PhotCalVers+ table 
(see Fig~\ref{fig:ERMs}a for the design of the recalibration procedure in the 
database).

Updates of the metadata are relatively efficient as the database tables
contain $\sim10^6$ rows and only the headers of the FITS files
need to be modified. Recalibration of WSA metadata takes one hour for every
five thousand detector images and for UKIDSS DR2 took just over one day. 
However, updates to the catalogue data in the database detection
tables is very slow as these tables can contain $\sim10^9$ rows. For the WSA, 
the
table of sources merged across passbands and epochs would also need to be
either entirely recreated or updated to reflect new calibration. For the
UKIDSS DR2 recalibration, all detections from the same detector image
required the same adjustment to the calibrated quantities. This translates
to a reasonably efficient SQL update statement, and so recalibration of
detection data was on average as efficient as the metadata and took one
additional day for the largest table. However, the merged
\verb+Source+ table still had to be recreated from scratch, which takes two 
weeks for the largest table.

Future recalibrations will be less efficient. For example
for UKIDSS DR4, there will be three changes to the calibration where 
the correction will vary across the detector such that each row must have 
an independent update. One is a change to the astrometric solution where 
low-level 2-dimensional systematic effects across each detector have been 
found. Similar improvements to the photometric solution will also be put in 
place, including an improvement to the pixel distortion correction which adds 
an extra radial term. These recalibration updates will be at least a factor 
of ten slower, and so will take more than a month to perform at the current 
data volume. Therefore, a new, more efficient approach to the database 
recalibration problem is required.

One solution to the problem of inefficient SQL updates is to minimise the
time lost to SQL statement compilation and transaction logging by
recalibrating the data outside of the database and then bulk load the data
into a table. To reingest into the WSA $10^9$ rows of detection data would
take several weeks (see Sutorius et al. 2008). Although faster than the
current method, this is still impractical for the data volumes that will be
stored in the VSA.

\vspace{-5mm}
\section{A high-performance database recalibration design}

To improve the performance of the recalibration of the database, 
we have chosen to redesign the database schema to optimise curation 
performance, whilst maintaining the original schema in the 
release database that we present to our users. Recalibration is most 
efficient when the calculations are performed outside of the database, with 
the recalibrated data then being bulk loaded back into the database table. 
To optimise this procedure we have chosen to denormalise the database schema 
design for the \verb+Detection+ tables, breaking them up into three 
separate tables; one containing raw uncalibrated quantities, one containing 
the calibrated photometric quantities and one containing the calibrated 
astrometric quantities, all joined by the primary key in a one-to-one 
relationship. Therefore, whenever recalibration occurs, only the values that
change need to be recalculated and bulk loaded back into the database,
minimising the data flow. Furthermore, a new merged source table, 
\verb+mergeSource+, is introduced for curation that contains the attributes
of the Source table minus the calibrated photometric quantities: 
avoiding the need to update or recreate this table following a photometric 
recalibration event. At release, the three 
denormalised detection tables are joined, and the calibrated quantities 
inserted back into the released merged \verb+Source+ table. This maintains 
schema 
consistency in our released databases, whilst improving the performance 
of data curation.

Our new database design is shown in Fig~\ref{fig:ERMs}b. For the sake of 
clarity, we have left out the \verb+PreviousMFDZP+, \verb+PhotCalVers+, and 
equivalent astrometry tables. Recalibration now consists of recalculating 
either the astrometric or the photometric calibrated quantities (or both, 
which may be performed in parallel). The relevant table(s), e.g. 
\verb+DetectionPhotometry+, 
\verb+DetectionAstrometry+, are then dropped, recreated and bulk loaded 
with the newly calculated derived quantities. Reducing the
width of the database table that requires photometric
recalibration by a factor of three, and astrometric recalibration by a
factor of six, gives corresponding improvements in the time required to bulk
load the data. Time is also not wasted in recalculating derived quantities
that have not changed. By shrinking the merged source table down to
a third of its previous width, the performance of the source merging process
will also be improved.

\end{document}